\documentclass[10pt, conference, compsocconf]{IEEEtran}
\IEEEoverridecommandlockouts
\usepackage{cite}
\usepackage{amsmath,amssymb,amsfonts}
\usepackage{algorithmic}
\usepackage{graphicx}
\usepackage{textcomp}
\usepackage{xcolor}
\usepackage{amsmath, bm}
\usepackage{multicol, multirow}
\usepackage{subfig}
\usepackage[normalem]{ulem}
\useunder{\uline}{\ul}{}
\usepackage[switch]{lineno}
\usepackage{wrapfig,lipsum,booktabs}
\usepackage{setspace}
\usepackage{wrapfig}
\usepackage{fnpos} 

\usepackage[ruled,vlined,linesnumbered,noresetcount]{algorithm2e}
\makeatletter
\newcommand{\AlgoResetCount}{\renewcommand{\@ResetCounterIfNeeded}{\setcounter{AlgoLine}{0}}}
\newcommand{\AlgoNoResetCount}{\renewcommand{\@ResetCounterIfNeeded}{}}
\newcounter{AlgoSavedLineCount}

\makeatother

\def\BibTeX{{\rm B\kern-.05em{\sc i\kern-.025em b}\kern-.08em
    T\kern-.1667em\lower.7ex\hbox{E}\kern-.125emX}}

\begin{document}

\title{Preliminary Investigation of SSL for Complex Work Activity Recognition in Industrial Domain via MoIL}

\author{\IEEEauthorblockN{Qingxin Xia}
\IEEEauthorblockA{\textit{Multimedia Engineering} \\
\textit{Osaka University, JP}\\
xia.qingxin@ist.osaka-u.ac.jp}
\and
\IEEEauthorblockN{Takuya Maekawa}
\IEEEauthorblockA{\textit{Multimedia Engineering} \\
\textit{Osaka University, JP}\\
takuya.maekawa@acm.org}
\and
\IEEEauthorblockN{Jaime Morales}
\IEEEauthorblockA{\textit{Multimedia Engineering} \\
\textit{Osaka University, JP}\\
jaime.morales@ist.osaka-u.ac.jp}
\and
\IEEEauthorblockN{Takahiro Hara}
\IEEEauthorblockA{\textit{Multimedia Engineering} \\
\textit{Osaka University, JP}\\
hara@ist.osaka-u.ac.jp}
\and
\IEEEauthorblockN{Hirotomo Oshima}
\IEEEauthorblockA{\textit{Corporate Manufacturing Engineering}\\
\textit{Center, Toshiba Corporation}\\
hirotomo1.oshima@toshiba.co.jp}
\and
\IEEEauthorblockN{Masamitsu Fukuda}
\IEEEauthorblockA{\textit{Corporate Manufacturing Engineering}\\
\textit{Center, Toshiba Corporation}\\
masamitsu1.fukuda@toshiba.co.jp}
\and
\IEEEauthorblockN{Yasuo Namioka}
\IEEEauthorblockA{\textit{Corporate Manufacturing Engineering}\\
\textit{Center, Toshiba Corporation}\\
yasuo.namioka@toshiba.co.jp}
}

\maketitle

\begin{abstract}
\textcolor{black}{In this study, we investigate} a new self-supervised learning (SSL) approach for complex work activity recognition using wearable sensors. 
Owing to the cost of labeled sensor data collection, SSL methods for human activity recognition (HAR) that effectively use unlabeled data for pretraining have attracted attention. 
However, applying prior SSL to complex work activities such as packaging works is challenging because the observed data vary considerably depending on situations such as the number of items to pack and the size of the items in the case of packaging works.
In this study, we focus on sensor data corresponding to characteristic and necessary actions (sensor data motifs) in a specific activity such as a stretching packing tape action in an assembling a box activity, and \textcolor{black}{try} to train a neural network in self-supervised learning so that it identifies occurrences of the characteristic actions, i.e., Motif Identification Learning (MoIL). 
The feature extractor in the network is used in the downstream task, i.e., work activity recognition, enabling precise activity recognition containing characteristic actions with limited labeled training data. 
The MoIL approach was evaluated on real-world work activity data and it achieved state-of-the-art performance under limited training labels.

\end{abstract}


\section{Introduction}

\begin{figure}[t!]
  \begin{center}
\includegraphics[width=0.95\linewidth]{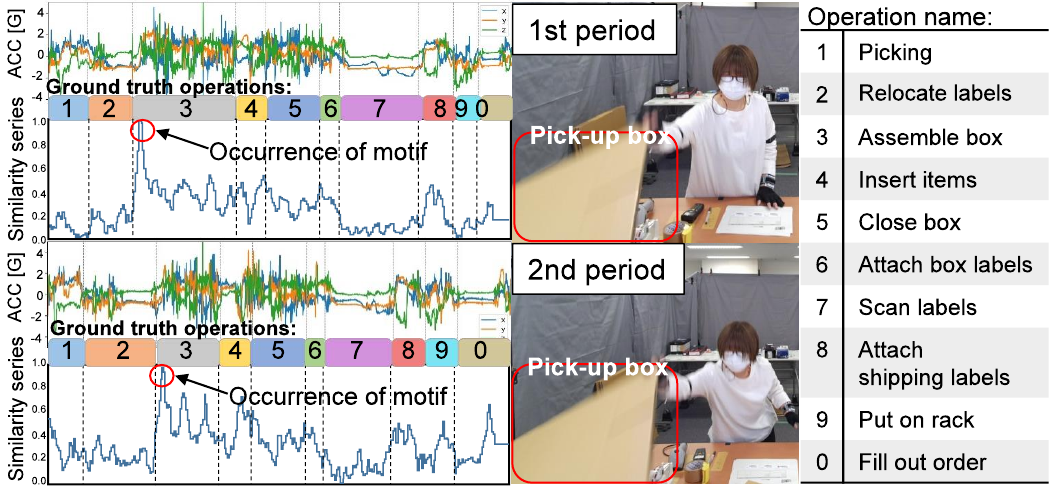}
\caption{Example of an occurrence of a motif corresponds to the pick-up box action in two periods.}
\label{fig:example}
  \end{center}
\end{figure}

\textbf{Background}.
In industrial domains, human workers remain crucial for keeping quick responses to the rapidly changing demands of customers and suppliers, and the requirements on workers continue to grow in recent years \cite{info10080245}.
Commercial wearable devices have attracted attention for their effectiveness in detecting complex activities,
thus, studies on HAR for human workers using wearable sensors have become prevalent in recent years in the pervasive computing community \cite{Jaime, yoshimura, docxia}. 
Figure \ref{fig:example} shows a typical example of packaging work with complex activities. The worker performs a set of operations (activities) repetitively, with each operation consisting of a sequence of small actions. For example, an operation of ``assemble box'' consists of atomic actions such as folding all sides of the box, stretching a packing tape, 
\textcolor{black}{and etc.}
A complete packaging task is called a period.
The objective of the worker activity recognition task in industrial settings is to estimate the class of operations (activities) for each time step (data point).


\textbf{Challenges}.
Different from daily life activity recognition, recognizing complex activities in industrial settings has many challenges. 
(1) \textbf{Complex sensor data}. 
An operation in industrial domains is composed of different small actions rather than periodic patterns such as walking \cite{dailyactivitysurvey}.
(2) \textbf{Highly variable sensor data}. As shown in Figure \ref{fig:example}, the same operation such as ``assemble box'' differs in both periods in terms of duration and actions involved, due to the difference in items to pack. 
(3) \textbf{Limited training data}. Owing to challenges (1) and (2), training the HAR model for manual work requires a larger amount of labeled data, while collecting labeled data is costly. 
Recent works \cite{SSLsurvey,maskedReconstruct} apply SSL to daily life activity recognition to address the limited training data issue. However, state-of-the-art SSL methods \textcolor{black}{have led to a deterioration in their performance in complex activity recognition (refer to Section \ref{sec:eval})}. Thus, there is a demand for a novel SSL approach tailored to the activities in industrial domains, despite the complex and variable nature of the actions that constitute these activities.

\textbf{Approach}.
In this study, we \textcolor{black}{investigated a new} SSL approach, named Motif Identification Learning (MoIL).
As mentioned above, the operations involve several atomic actions, an atomic action exhibits a characteristic waveform in sensor data, called a motif. For example, the pick-up box action always occurs in the assembling box operation even though the packaging task is highly variable.
Automatically detecting such motifs in the acceleration data is useful for precise activity recognition. 
By designing an SSL pretext task to learn latent representation regarding the characteristic motifs, we can improve the performance of the downstream task (i.e., complex activity recognition) with a small amount of annotated data.
In this study, we perform MoIL to identify occurrences of motifs within an input time window of unlabeled acceleration data by reconstructing the \textbf{similarity series}, which shows the occurrences of the corresponding key action in the acceleration data. 
Thus, the network can be trained to detect key motifs that correspond to characteristic atomic actions. 
Therefore, the feature extraction layers of the network can be effectively used in the downstream task, i.e., recognizing complex operations consisting of multiple key actions.

\section{SSL by Motif Identification Learning (MoIL)}

Our methodology consists of three components: (1) Key motif selection, (2) MoIL for the pretext task, and (3) activity recognition in the downstream task. 
%
We collect labeled/unlabeled acceleration data from workers using a body-worn accelerometer in advance. 
The labeled/unlabeled data consists of a set of sensor data sequences of multiple work periods. 
\textcolor{black}{Here, a period of data is represented as $\bm{X} = [\bm{x}_1, \bm{x}_2, ..., \bm{x}_T]$, where $T$ represents the length of the period and $\bm{x}_t$ represents the sensor values at time step $t$. 
The sets of sensor data sequences for the unlabeled, labeled, and all work periods are denoted as $\bm{D}_u$, $\bm{D}_l$, and $\bm{D}_a$, respectively, $\bm{D}_a=\bm{D}_u\cup \bm{D}_l$.
}


\subsection{Key Motif Selection}

\subsubsection{Motif Candidates 
 Generation}\label{sec:standardize}
\textcolor{black}{Firstly, we apply min-max normalization and symbolization for raw acceleration data to efficiently find key motifs. For symbolization, we convert acceleration values belonging to the same range to the same symbols based on \cite{symbolization}.}
After symbolization, we generate \textcolor{black}{a set of motif candidates $\bm{M}$ from an initial period $\bm{X}$ ($\bm{X}$ is randomly selected from $\bm{D}_u$).}
We use a sliding window with a fixed step across the symbolized data to extract \textcolor{black}{data segments} as candidate motifs.
A group of candidates of different lengths is generated using different window sizes.

\subsubsection{Calculating Similarity Series}\label{sec:calculatesimi}

For each candidate motif $m$, we calculate the similarity series of $m$ over all unlabeled periods $\bm{D}_u$. 
\textcolor{black}{Let $\bm{X}^i$ be the $i$-th period of data in $\bm{D}_u$,} $\bm{S}_i^m$ be the similarity series of \textcolor{black}{period} $\bm{X}^i$ for $m$, and $g(\cdot)$ denote the symbolization function. $\bm{S}_i^m$ is formulated as:
\begin{equation}
\small
\begin{aligned}
    \bm{S}_i^m = [-d(m, g(\bm{X}^{i[1:1+|m|]})), -d(m,g(\bm{X}^{i[2:2+|m|]}))\\
    ,..., -d(m,g(\bm{X}^{i[|\bm{X}^{i}|-|m|:|\bm{X}^{i}|]}))] 
\end{aligned}, 
\end{equation}
where $|m|$ and $|\bm{X}^{i}|$ correspond to the length of $m$ and $\bm{X}^{i}$, respectively.
$d(,)$ represents a distance metric between two segments.
Inspired by the Hamming distance \cite{hamming}, we compute the distance between two symbolized segments by counting the number of positions at which the corresponding symbols are dis-similar, \textcolor{black}{enabling us to reveal the similarity of actions at each time step effectively.}
\textcolor{black}{Finally, we average $\bm{S}_i^m$ over all the axes and normalize the values to $[0,1]$.}

\subsubsection{Selecting Motifs among Candidates}
\textcolor{black}{Because if two motifs located too close in time, information obtained from the motifs will be similar. 
In this method, we evenly divide the initial period into $n$ segments ($n = 13$), and form a group consisting of candidate motifs extracted from each segment in the initial period. We randomly select a motif among each group as the key motif. Subsequently, the corresponding similarity series for the key motifs will be processed to MoIL.}

\subsection{MoIL}

\begin{figure}[t]
  \begin{center}
\includegraphics[width=0.84\linewidth]
{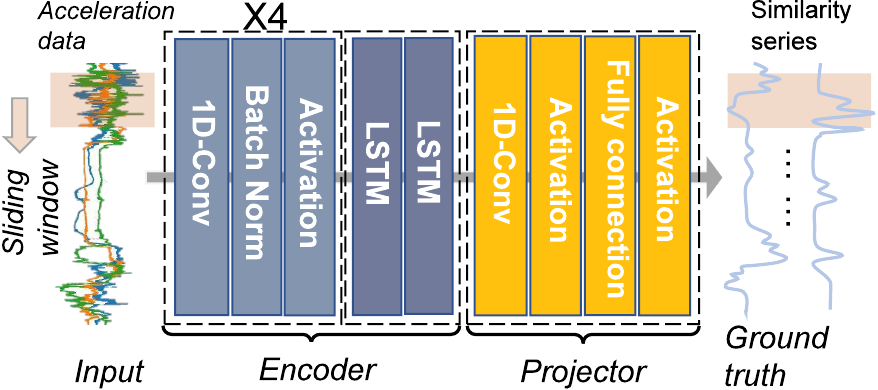}
\caption{Overview of MoIL.}
\label{fig:MoIL}
  \end{center}
\end{figure}

\textcolor{black}{Here we pretrains a feature extractor (encoder) for complex activity recognition via SSL. The trained feature extractor obtains features related to selected key motifs, which is helpful for the downstream task.}

\subsubsection{SSL Model}
As shown in Figure \ref{fig:MoIL}, MoIL is composed of an encoder and a projector. The input is a \textcolor{black}{sensor data segment}, and the ground truth is the corresponding similarity series \textcolor{black}{segment}. 
Note that since we have $n$ key motifs, the model outputs an ($n$)-dimensional similarity segment.

In this work, the encoder consists of four CNN blocks and two RNN blocks. 
The rationale for using the convolutional layers is \textcolor{black}{to extract the local temporal dependencies of the sensor data. The output of the $i$-th block is fed into the $i+1$-th block, which is described as follows:}
%
\begin{equation}
    \mathbf{X}_u^{B(i+1)} = \text{ReLU}(\text{BatchNorm}(\text{1DConv}(\mathbf{X}_u^{B(i)}))) ,
\end{equation}
\textcolor{black}{where $\mathbf{X}_u^{B(i)}$ is} the input of the $i$-th CNN block ($i=1,2,3,4$); $\text{1DConv}(\cdot)$ denotes the 1D convolutional layer \textcolor{black}{(64 kernels with a kernel size of 5 and a stride of 1)}; $\text{BatchNorm}(\cdot)$ and $\text{Sigmoid}(\cdot)$ denote the Batch Normalization and the Sigmoid activation function, respectively.
To extract long temporal dependencies of the sensor data, two RNN blocks are used. 
Each RNN block has a bidirectional LSTM layer with 128 units. The output of the last CNN block is fed into the first RNN block, and the output of the first RNN block is fed into the second RNN block, which is described as follows.
\begin{equation}
    \mathbf{X}_u^{B(i+1)} = \text{LSTM}(\mathbf{X}_u^{B(i)}) ,
\end{equation}
where $\text{LSTM}(\cdot)$ denotes the RNN block ($i=5,6$).

The projector adjusts the output shape of the encoder module to be suitable for reconstructing similarity series.
Herein, the projector is composed of a 1D convolutional layer with the same settings as the mentioned 1DConv($\cdot$); a ReLU activation function $\text{ReLU}(\cdot)$; a fully connected layer $\text{Linear}(\cdot)$ with the dimension of the output features of the layer equal to the number of similarity series for the key motifs; a ReLU activation function. The structure is described as follows:
\begin{equation}
    \mathbf{X}_u^{B(i+2)} = \text{ReLU}(\text{1DConv}(\mathbf{X}_u^{B(i+1)})) ,
\end{equation}
\begin{equation}
    \hat{\bm{S}}_N =  \text{ReLU}(\text{Linear}(\mathbf{X}_u^{B(i+2)})) ,
\end{equation}
where $\hat{\bm{S}}_N$ is \textcolor{black}{an estimates} of a $N$-dimensional \textcolor{black}{similarity series segment} ($N = n$).

\subsubsection{Network Training}
We employ the mean squared error (MSE) as the loss function for the SSL task in MoIL, which is shown as follows:
\begin{equation}
    L_{ssl} = \frac{1}{N} \sum_{j=1}^{N} \frac{1}{l} \sum_{i=t}^{t+l} (\hat{s}_{i,j} - s_{i,j})^2,
\end{equation}
where $\hat{s}_{i,j}$ and $s_{i,j}$ are the $i$-th prediction and ground truth values in the similarity series of the $j$-th key motif, respectively. \textcolor{black}{$l$ is the segment length.} Model parameters are optimized via the Adam optimizer \cite{adam}.

\subsection{Activity Recognition}
The neural network of the downstream task consists of an encoder directly copied from the SSL model and a classifier.
The classifier is an MLP module consisting of three linear layers of 256, 128, and $C$ units, where $C$ is the number of activity classes. \textcolor{black}{A Batch Normalization and activation function} are applied consecutively between each layer.
During the model training, we freeze the learned weights of the encoder module from MoIL and optimize the parameters of only the classifier. Given a \textcolor{black}{window of sensor data $\bm{X}^{[t:t+l]}$ as the input} and $\bm{Y}^{[t:t+l]}=[\bm{y}_t, \bm{y}_{t+1}, ..., \bm{y}_{t+l}]$ as the ground truth label, the classifier is trained to output estimates that have minimum errors to $\bm{Y}^{[t:t+l]}$. 
We train the downstream model using the cross-entropy loss using the Adam optimizer as follows: 
\begin{equation}
    L_{ar} = - {\textstyle  {\sum_{i=t}^{t+l}} \sum_{c=1}^{C}} \bm{y}_{ic} \text{log}(p_{ic}),
\end{equation}
\textcolor{black}{where $\bm{y}_{ic}$ is a one-hot vector corresponding to the $i$-th prediction of class $c$,}
and $p_{ic}$ represents the prediction \textcolor{black}{of $x_l^i$ belonging} to class $c$.
The idea of training the classifier only is to compare the performance of feature extraction via MoIL against other state-of-the-art SSL frameworks when using limited data annotations.

\section{Evaluation}\label{sec:eval}

\subsection{Datasets}


\subsubsection{OpenPack Dataset\cite{openpack}} 
The workers were asked to repeatedly perform a packaging task several times by following the work instruction document (Scenario 1, \textcolor{black}{all 10 workers}). 
Acceleration data from their both wrists were collected using Empatica E4 wristband with a sampling rate of 30Hz.


\subsubsection{Logi Dataset} 
This private dataset contains workers performing packaging tasks in a real logistics center.
Data is collected using a smartwatch worn on the workers' dominant hands (Sony SmartWatch3 SWR50) with a sampling rate of 30 Hz.
\textcolor{black}{Unlike the OpenPack dataset, the order and number of operations in the Logi dataset vary for each period.}

\subsubsection{TestBoard Dataset} 
\textcolor{black}{This private dataset contains a worker performing testing circuit boards task in a factory using a smartwatch worn on the dominant hand (Sony SmartWatch3 SWR50) with a sampling rate of 30 Hz.}

\subsubsection{Skoda Dataset \cite{skoda}} A single worker working on an assembly line in car manufacturing.
\textcolor{black}{We process the dataset to form periods of data following the approach used in \cite{yoshimura}.}

\subsection{Evaluation \textcolor{black}{Methodology}}

A sliding window was applied \textcolor{black}{acceleration data} to generate segments of time-series data, which were inputs of the pretext and downstream tasks. The window size was set to 30 seconds (i.e., 900 data points).
\textcolor{black}{For the training set, the step of the sliding window was 15 seconds, while for the test set, the step was 30 seconds.}
The activity recognition accuracy was calculated as the micro average F1-measure of all the time steps (i.e., data points) in the test set. 
We conducted five times experiments by changing the random seed, and the standard deviation was calculated according to the F1-measure of the five experiments. 
The hyper-parameters are shown in Table \ref{tab:experimentparam}.

\begin{table}[]
\footnotesize
\caption{Experimental parameters.}
\begin{tabular}{l|l}
\hline
\textbf{Model training} &
  \begin{tabular}[c]{@{}l@{}}SSL learning rate: 1e-4, classifier learning rate: 1e-3, \\ L2 regression: 1e-4, batch size: 1000,\\ SSL \#.epochs: 1000, downstream training \#.epochs: 50\end{tabular} \\ \hline
\end{tabular}
\label{tab:experimentparam}
\end{table}

\subsection{Baselines}
\noindent
\textbf{Multi-task self-supervision \cite{sslMultitask}}. This is a multi-task SSL approach that applies eight signal transformations and assigns a binary classification task for each transformation, i.e., transformed or not. \\
\textbf{Masked reconstruction \cite{maskedReconstruct}}. This task is similar to the \textcolor{black}{BERT} model, which learns latent representation by reconstructing the masked input signals. \\
\textbf{Autoencoder \cite{autoencoder}}. This task contains an ``encoder-decoder'' structure, which aims to reconstruct the original signal using the MSE loss. \\ 
\textbf{SimCLR, BYOL, and SimSiam}. \textcolor{black}{They are the state-of-the-art contrastive learning models for SSL. We follow the experiment settings the same as \cite{SSLsurvey}.
}


\subsection{Results}
\subsubsection{Worker-dependent Model}
\textcolor{black}{This experiment measures the effectiveness of MoIL in feature extraction on complex activities compared to other state-of-the-art SSL methods.}
\textcolor{black}{Same experiment setting as used in the survey paper \cite{SSLsurvey},} we divided all the periods collected from each worker into training and test sets, with \textcolor{black}{the first 80\% of periods for training and the remaining 20\% of periods for test}, 
and \textcolor{black}{only} the training set was used for pretraining in the SSL as unlabeled data.

\begin{figure}
\centering
\includegraphics[scale=0.35]
{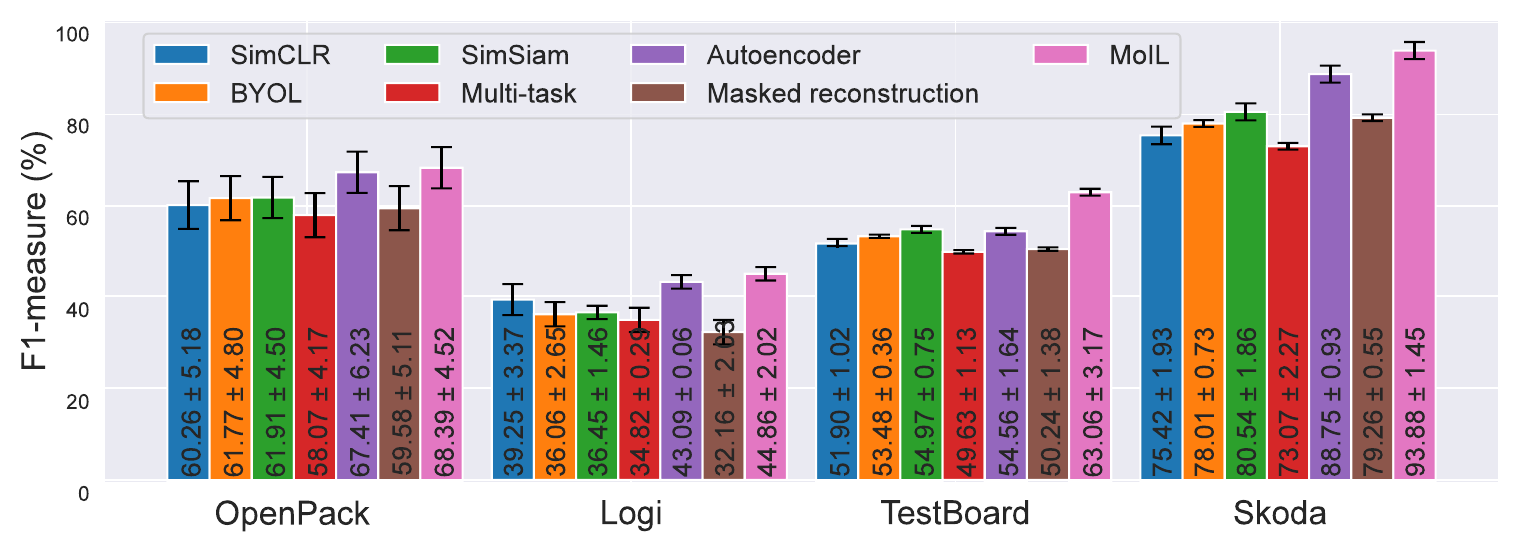}
\caption{F1-measure (\%) of the SSL methods for each dataset.}
\label{fig:resultwip}
\end{figure}

Figure \ref{fig:resultwip} shows the average F1-measure for each SSL method.
MoIL outperformed the other SSL methods among all datasets, indicating that MoIL can effectively learn latent representation for activity recognition using unlabeled data despite the variation of workers and tasks.
\textcolor{black}{In particular, the factory work tasks (TestBoard and Skoda) showed better performance than the packaging tasks. This could be because the operations performed in the factories are more consistent and the activities in different operations are similar (raw sensor data example of TestBoard can be found in \cite{docxia}), thus the similarity series, which calculates the similarities between actions, can effectively identify the difference among operations.}
Although the F1-measure on the Logi dataset was relatively low because the occurrence and duration of operations greatly differ in every period \textcolor{black}{(refer to Table III of DeepConvLSTM in \cite{Jaime}, even using more than 20\% of labels, the F1-measure of the supervised learning method could not reach 60\%)}, features extracted by MoIL are more effective than the other SSLs.

\begin{figure}
\centering
\includegraphics[scale=0.77]
{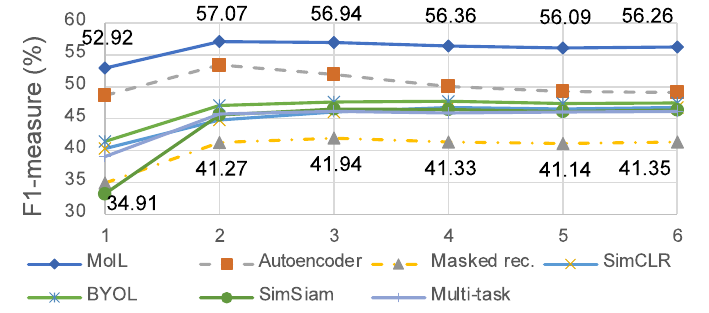}
\caption{Average F1-measure (\%) of leave-one-worker-out for OpenPack dataset \textcolor{black}{(worker-independent models)}.}
\label{fig:leaveoneworker}
\end{figure}

\subsubsection{\textcolor{black}{Worker-independent Model}} 
This experiment investigates if the latent representation learned from other workers is helpful for a new worker. \textcolor{black}{Note that Workers who appear in the test set will not have any data used in the training set.}
Taking the OpenPack dataset as an example, we first utilized the other workers' entire unlabeled data of the OpenPack dataset for pretraining, then we fixed the parameters of the encoder and used their labels to train a new classifier connecting to the encoder. \textcolor{black}{Finally,} we applied the trained model to predict the class label of \textcolor{black}{the} new worker in OpenPack.  
\textcolor{black}{For the downstream task, we recorded the F1-measure testing on every new worker when the training iteration that uses the other workers' labels reached 1, 10, 20, 30, 40, and 50, and Figure  \ref{fig:leaveoneworker} illustrates the average F1-measure for the new workers in the OpenPack dataset.}
\textcolor{black}{Overall, MoIL demonstrated superior performance compared to the other SSL baselines. This result suggests that latent representations pretrained using MoIL on other workers' data can be successfully applied to a new worker, which may be because the reconstruction of similarity series could help the model learn the characteristics of actions, \textcolor{black}{highlighted MoIL's proficiency in complex activity recognition}. Besides, MoIL exhibited a trend of initially increasing and then decreasing in F1-measures. This behavior may be attributed to the variations in actions involved in operations among different workers. Training a classifier using operation labels from other workers could potentially deteriorate the classification performance.}

\section{Conclusion}
We \textcolor{black}{investigated} a new SSL approach MoIL for sensor data representation learning focusing on complex work activity recognition in the industrial domain.
We evaluated our approach and demonstrated that MoIL outperformed state-of-the-art SSL baselines on four industrial tasks.
As part of our future work, we plan to investigate various motif selection strategies.

\section*{Acknowledgement}
This research is partially supported by JST CREST Grant Number JPMJCR21F2.
\bibliographystyle{IEEEtran}
\bibliography{reference}

\end{document}